# Superconductivity in the surface state of noble metal gold and its Fermi level tuning by EuS dielectric

**Authors:** Peng Wei[1,3†*], Sujit Manna[1†*], Marius Eich[2,4], Patrick Lee[1*] and Jagadeesh Moodera[1,2*]

**Affiliations:**

1. Department of Physics, Massachusetts Institute of Technology, Cambridge, MA 02139, United States
2. Plasma Science and Fusion Center & Francis Bitter Magnet Laboratory, Massachusetts Institute of Technology, Cambridge, MA 02139, United States
3. Department of Physics and Astronomy, University of California, Riverside, CA 92521, United States
4. Solid State Physics Laboratory, ETH Zurich, 8093 Zurich, Switzerland

† These authors contributed equally to this work

* Correspondence should be addressed to:

  peng.wei@ucr.edu

  smanna@mit.edu

  palee@mit.edu

  moodera@mit.edu

**Abstract:**

The induced superconductivity (SC) in a robust and scalable quantum material with strong Rashba spin-orbit coupling is particularly attractive for generating topological superconductivity and Majorana bound states (MBS). Gold (111) thin film has been proposed as a promising candidate because of the large Rashba energy, the predicted topological nature and the possibility for large-scale MBS device fabrications.[1] We experimentally demonstrate two important steps towards achieving such a goal. We successfully show induced SC in the Shockley surface state (SS) of ultrathin Au(111) layers grown over epitaxial vanadium films, which is easily achievable on a wafer scale. The emergence of SC in the SS, which is physically separated from a bulk superconductor, is attained by indirect quasiparticle scattering processes instead of by conventional interfacial Andreev reflections.[1] We further show the ability to tune the SS Fermi level ($E_F$) by interfacing SS with a high-κ dielectric ferromagnetic insulator EuS. The shift of $E_F$ from ~ 550 mV to ~34mV in superconducting SS is an important step towards realizing MBS in this robust system.

**One Sentence Summary:** Surface state superconductivity near the Kramer's degeneracy point of Rashba band in ultra-thin gold with strong spin-orbit coupling.

**Main text:**

The Shockley surface state (SS) of Au(111) is well-known to feature strong Rashba spin-orbit coupling (SOC) and recently has been predicted to have topological nature.[2,3] Such strong SOC, reaching 110 meV and orders of magnitude stronger than those in semiconductors, has been theoretically shown to produce robust MBS once the SS attains SC.[1] Experimentally achieving SC in SS of Au, therefore, will lay the foundation for obtaining much more robust MBS. Since large-scale design and fabrication of nanowire network out of wafer-scale Au(111) thin film is possible, Au(111) is a scalable platform that would realize a variety of proposed schemes to manipulate the MBS.

The SS band has no direct overlap with the projection of the bulk gold band on the (111) surface and the standard mechanism of superconducting proximity effect does not apply. Our prior theoretical work has shown that a finite superconducting pair amplitude is generated in SS through elastic or inelastic scattering processes even though the SS are not directly in touch with a superconductor.[1] Although it is not clear how this new mechanism will work in practice, it is important to demonstrate the superconductivity in SS experimentally. A second problem is that the Fermi energy of the SS is very large, ~550 meV, and holds many transverse sub-bands for any realistic nanowire width. It will be important to reduce the Fermi energy. In this paper we demonstrate that by depositing EuS on the gold surface, a giant shift from ~550 meV to ~34 meV is achieved. Further, EuS has the added advantage that it is a ferromagnet insulator and can enhance the Zeeman energy by exchange interaction. By using EuS as a barrier in tunnel junction, we see evidence of the magnetism by change of coherence peak height as a function of the magnetic field.

We have achieved surface state SC by using pristine (111) surface of ultra-thin (4 nm) wafer-scale Au film grown on vanadium, a *s*-wave superconductor.[4] The film thickness ~ 4 nm is chosen so that the Au(111) layer thickness is much greater than the SS penetration depth (~3.2 monolayer),[5] and thin enough to allow fully induced SC gap in its bulk,[6] in contrast to the previous report on Ag(111) where the SS penetration depth is comparable to the layer thickness.[5,7] Our island-free Au(111) layer on V (Fig. 1) further allows uniformly induced SC gap, which could otherwise be degraded by island boundaries,[7-9] providing a stable platform for fabricating scalable Au nanowire network in the future for detecting and braiding the MBS.[1,10]

The high quality Au(111) layer, confirmed by atomically resolved STM topography (Fig. 1c and 1d), possesses clear SS. The bottom of the SS band ($E_{SS}$) manifests as a peak in $dI/dV$ vs $V_{bias}$ spectrum at $V_{bias}$ ~ -0.57 eV (Fig. 1c inset and Fig. 2b). Compared to bulk Au crystals,[11,12] our measured $E_{SS}$ is close but lower. This could be a result of the STM tip electric field affecting the SS, or could indicate that the value of $E_{SS}$ may be different in 4nm Au(111) grown on V. At $V_{bias}$ = 0 eV, where $E_F$ is located, the Fermi level therefore crosses both bulk states and SS. It has been shown that the SS of a noble metal can degrade when growing on another metal and having a thickness comparable to the SS decaying depth.[5] The sharp $dI/dV$ peak (Fig. 1c inset) demonstrates that the SS is well defined.

We show that $E_{SS} - E_F$ sensitively depends on the dielectric environment above the Au(111) surface (Fig. 2), an effect which is crucial for electronically manipulating MBS in future experiments.[1,12] We modulate $E_F$ by growing ultra-thin EuS, a ferromagnetic high-$\kappa$ dielectric insulator ($\epsilon$ ~ 23.9 in bulk),[13] on top of Au(111)/V (Fig. 2a). Rectangular shaped EuS islands with uniform monolayer (ML) height (~2.8 Å) grow well on Au(111). In order to estimate how a ML of EuS affects the surface state of the underneath Au, we measure spectroscopy on top of EuS island and bare Au surface sequentially (see Fig. 2b). On both bare and EuS covered sites, $dI/dV$ shows a peak at $E_{SS}$ (Fig. 2b) as expected. Compared to bare Au, $E_{SS}$ on EuS island is shifted upwards by ~200 meV towards $E_F$. We point out that the Au(111) surface has the same $E_F$ regardless of whether it is with or without the coverage of EuS, whereas the reduced $|E_{SS} - E_F|$ is a result of the increase of $E_{SS}$ due to EuS coverage. Because the SS are quantum well states confined by the *s-p* bulk band gap of Au and the surface image potential, the large dielectric constant of EuS modifies the image potential and reduces the quantum well width to deplete the surface electrons.[14] Unlike other reported approaches of tuning the SS band of Au(111), for example using monolayer MgO,[15] the magnetic EuS also generates a substantial interface Zeeman field (ZF), a prerequisite for creating MBS in Au(111).[1,16-19]

Next, we bring $E_F$ further closer to the Kramers degeneracy point of SS by increasing EuS thickness. With thicker EuS layer scanning tunneling spectroscopy (STS) becomes impractical, so we fabricated planar thin film sandwich tunnel junctions (TJs) and perform $dI/dV$ tunneling spectroscopy through the 2.4 nm thick EuS layer (Fig. 2c). In TJs, we observe that $E_F$ approaches $E_{SS}$ further with $|E_{SS} - E_F|$ ~ 34 meV (Fig. 2c), which is in the vicinity of the Kramers degeneracy point.[20] A sizable Zeeman field, such as that provided by the EuS layer,[16-18,21,22] is predicted to cause topological SC and MBS.[1] The depletion of the SS also serves to reduce the number of the SS sub-bands, leading to a much more favorable condition for the creation of MBS's. [1]

The induced bulk SC gap in the 4 nm Au(111) is expected to be governed by conventional proximity effect. According to McMillian's model,[6] when a normal metal is in contact with a superconductor, the normal metal obtains a proximity induced self-energy of:

$$\Delta_N = \frac{(\Gamma_S \Delta_N^{ph} + \Gamma_N \Delta_S^{ph})}{(\Gamma_S + \Gamma_N)} \quad (1)$$

where $\Delta_N^{ph}$ and $\Delta_S^{ph}$ denote the self-energies of the normal metal and the superconductor due to phonons ($\Delta_N^{ph} \sim 0$ for Au).[6] The energy scales $\Gamma_N$ and $\Gamma_S$ are defined as $\Gamma_N = \frac{\hbar}{\tau_N}$ and $\Gamma_S = \frac{\hbar}{\tau_S}$, where relaxation times $\tau_N$ and $\tau_S$ represent the time a quasiparticle spent in the Au(111) layer and the V layer respectively.[6] For films with a thickness $d$ less than the mean-free-path $l$, we have $\tau \sim \frac{d}{v_F}$. Taking the literature values $v_F \sim 8 \times 10^5$ m/s in Au and $v_F \sim 1.8 \times 10^5$ m/s in V,[20,23] we find $\Gamma_N \gg \Gamma_S$ for 4nm Au(111) grown on 20nm V. Therefore, we expect $\Delta_N \approx \Delta_S^{ph}$ (Eq. 1) indicating that the bulk states of Au(111) inherit the full SC gap from V. Such conventional proximity effect is revealed by our TJs with EuS barrier, in which a SC gap shows up when T is below the $T_C$ of Au(111)/V (Fig. 3a).[4]

As the temperature of the sample being lowered below 2.5 K, the $dI/dV$ coherence peaks split (Fig. 3a). We attribute such splitting feature to indicate the emergence of a new SC gap in Au(111), and that it does not correspond to the lifted spin degeneracy in Au(111) under the magnetic exchange field (MEF) of EuS as previously seen in Al.[21] In materials with strong SOC, such as Au(111), spin is not a good quantum number and thus the spin splitting of the quasiparticle density-of-states (DOS) is suppressed in the presence of strong MEF, as has been demonstrated in Al TJs with Pt scatters.[24]

To confirm that the spin splitting is suppressed, the Al2O3/Al interface in a standard Al/Al2O3/Al/EuS junction is decorated with a sub-monolayer of Au (Fig. 3b and SI). The typical $dI/dV$ spectra (Fig. 3b) show collapsed spin-split coherence peaks in the presence of only 0.6 Å Au at the interface. Using Maki-Fulde model,[25-28] we show that the spin-orbit scattering parameter $b$ (= $\hbar/3\Delta\tau_{so}$, where $\tau_{so}$ is the spin-orbit scattering time and $\Delta$ is the superconductor gap), systematically increases when the Au thickness increases (SI) and the split coherence peaks in 4nm Au (Fig. 3a) could not be caused by MEF. Hence the development of the second peak in Fig. 3a is attributed to the emergence of an additional SC gap that is opened up in Au(111) at temperatures below ~ 3K, lower than the bulk $T_C$ ~ 4.0 K of Au(111)/V.

To better resolve the emergent SC gap, the sample is cooled down to T = 1.0 K (Fig. 3c) and we observe four $dI/dV$ peaks labeled as $B_+$, $S_+$, $B_-$ and $S_-$ representing the sum (+) and difference (-) tunneling processes in a S-I-S TJ. The $dI/dV$ spectrum can be modeled by considering two SC gaps in the Au(111) layer: bulk gap (B) and surface gap (S) (Fig. 3d). The total quasiparticle DOS in Au(111) can be written as the sum of two BCS terms, i.e. one from bulk (B) and one from surface (S), as:

$$\left(\frac{1+r}{2}\right) \cdot Re\left(\frac{E-i\Gamma_B}{\sqrt{(E-i\Gamma_B)^2-\Delta_B^2}}\right) + \left(\frac{1-r}{2}\right) \cdot Re\left(\frac{E-i\Gamma_S}{\sqrt{(E-i\Gamma_S)^2-\Delta_S^2}}\right) \qquad (2)$$

with $r$ adjusting the ratio of the two, and $\Gamma$ the lifetime of the quasiparticles.[29] The model (Eq. 2) reproduces the position and the relative magnitudes of the $dI/dV$ peaks (Fig. 3c). We thereby extract the SC gaps as: $\Delta_B$ = 0.63 ± 0.04 meV, $\Delta_{Al}$ = 0.22 ± 0.04 meV, and $\Delta_S$ = 0.38 ± 0.02 meV. Both the bulk gap ($\Delta_B$) and the Al gap ($\Delta_{Al}$), as expected, nicely agree with the BCS relation $2\Delta = 3.5k_BT_C$ (Au/V $T_C$ ~ 4.0 K and Al $T_C$ ~ 1.7 K). The bulk gap $\Delta_B$ seen in Fig 3a originates from the induced bulk SC due to the conventional proximity effect as describe by Eq. 1. On the other hand, $\Delta_S$, with a smaller size, has to come from a different energy band in Au (111). Since $E_F$ crosses both bulk and surface bands (Fig. 2c and 2d), $\Delta_S$ is a result of the induced SC in the SS of Au(111). As discussed previously, the SS has a penetration depth of only ~ 3.2 monolayers in Au(111),[5] and is thus well separated from V by the 4nm thick Au(111) layer. Conventional proximity effect (Eq. 1) cannot account for this. Furthermore, the SS band lies within a large gap of the bulk bands when projected to the (111) surface which is responsible for the well-defined nature of the SS. This also means that direct single electron hopping is not possible between the surface and the bulk. Thus, the gap $\Delta_S$ must be induced indirectly – concurring with our previous theoretical predictions.[1] The idea is that while a single electron in the SS has no overlap with the bulk electronic state, a pair of SS electrons can couple to a Cooper pair in the bulk via elastic scattering from impurities or inelastic scattering due to phonons or Coulomb interactions. We point out that the peaks $S_+$, $B_-$ and $S_-$ do not result from the sub-gap bound states caused by Andreev reflections as explained below. Such bound states would satisfy the De Gennes - St. James equation as: $\frac{\varepsilon_n L}{\hbar v_F} = n\pi + \cos^{-1}\left(\frac{\varepsilon_n}{\Delta}\right)$ with $\varepsilon_n$ the energy of the bound states, $L$ the trajectory length of the coherent Andreev pairs, and $\Delta$ the main SC gap ($\Delta = \Delta_B$ in our case).[30] However, if we assume $S_-$, the peak positions of other sub-gap peaks ($S_+$ and $B_-$) cannot be reproduced by the De Gennes - St. James equation. Moreover, as shown in Fig. 3c, the tunneling conductance at $B_+$ is at least four times larger than those of the other peaks. This excludes also the possibility that they are contributed by the spin-split coherence peaks of Al under the MEF of EuS.[24] Therefore, the emergence of $\Delta_S$ is clearly attributable to the induced SC in the SS of Au(111).

The SC in SS further demonstrates contrasting properties under an external magnetic field suggesting its 2D nature. We apply a magnetic field $\boldsymbol{B}_\parallel$ of 270 Oe parallel to the film plane of the junction (Fig. 4a), large enough to align EuS magnetization.[19] The peak heights at $B_+$ and $B_-$ drop noticeably under $\boldsymbol{B}_\parallel$ (Fig. 4b) as a result of the orbital depairing effects. Because $B_+$ and $B_-$ correspond to the quasiparticle tunneling between the bulk states of Au(111) and Al, the enhanced depairing reflects the weakening of bulk SC.[24] However, the $S_+$ and $S_-$ peaks are noticeably increased (Fig. 4b) suggesting an improved quasiparticle lifetime, and which cannot be attributed to the tunneling of bulk states with $B_\parallel$ present. However, it is known that the depairing effect can be largely suppressed in superconductors approaching the 2D limit, where the thicknesses of the superconductor is smaller than the penetration depth of a parallel magnetic field.[24] Therefore, $S_+$ and $S_-$ could involve the tunneling of 2D-like quasiparticles such as those from the SS of Au(111).[5] Moreover, the field $B_\parallel$ aligns the magnetization of EuS. Before the application of $B_\parallel$, the non-aligned spins at the EuS/Au(111) interface and domain walls present in EuS can introduce spin-flip scattering, which reduces the quasiparticle lifetime of SS. The enhanced $S_+$ and $S_-$ peaks suggest the aligned interface spins and the magnetized EuS, further supporting the 2D nature of SS.

Interestingly, we observe an additional $dI/dV$ peak at zero voltage (Fig. 3c), which doesn't decay when $B_\parallel$ is applied, while it disappears when $B_\perp$ is applied. Our model in Eq. (2) cannot account for this $dI/dV$ peak. We suggest that this $dI/dV$ peak could be due to Andreev reflections, which cause the tunneling of Cooper pairs between the top electrode Al and the bottom electrode Au. Nevertheless, the physical origin of this $dI/dV$ peak requires further studies. Our demonstrated SC in the Au(111) SS sets the stage for the observation of MBS in tunable and scalable EuS/Au(111)/V layered system.

**Methods:**

We use similar growth method as reported before.[4] Scanning tunneling spectroscopy (STS) and scanning electron microscopy (SEM) are used to confirm the high quality Au(111) surface. The STM and STS experiments are performed in a custom assembled STM with RHK PanScan head integrated in a Janis 300mK $He^3$ cryostat with vector magnet. Atomically resolved

STM images of Au(111) surface are obtained after the growth by careful pumping of the sample space in the STM load-lock with a turbo molecular pump. The spectroscopy of differential conductance dI/dV versus bias voltage is performed at T = 5K under open feedback loop condition with a voltage modulation $V_{rms}$ ~ 7-10 mV and AC frequency f = 1.57 KHz.

**Figures and Captions:**

**Fig. 1**

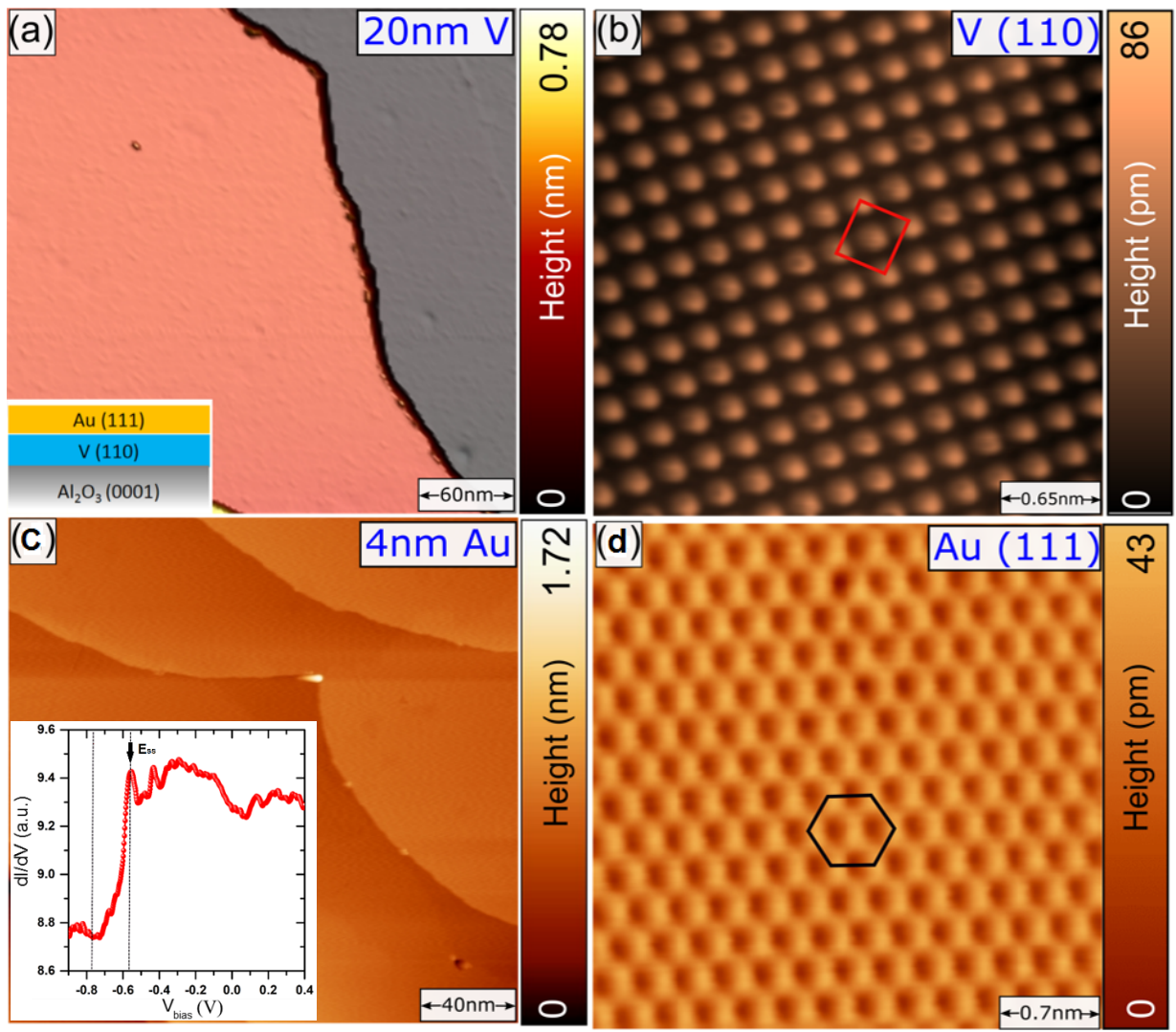

**Fig. 2**

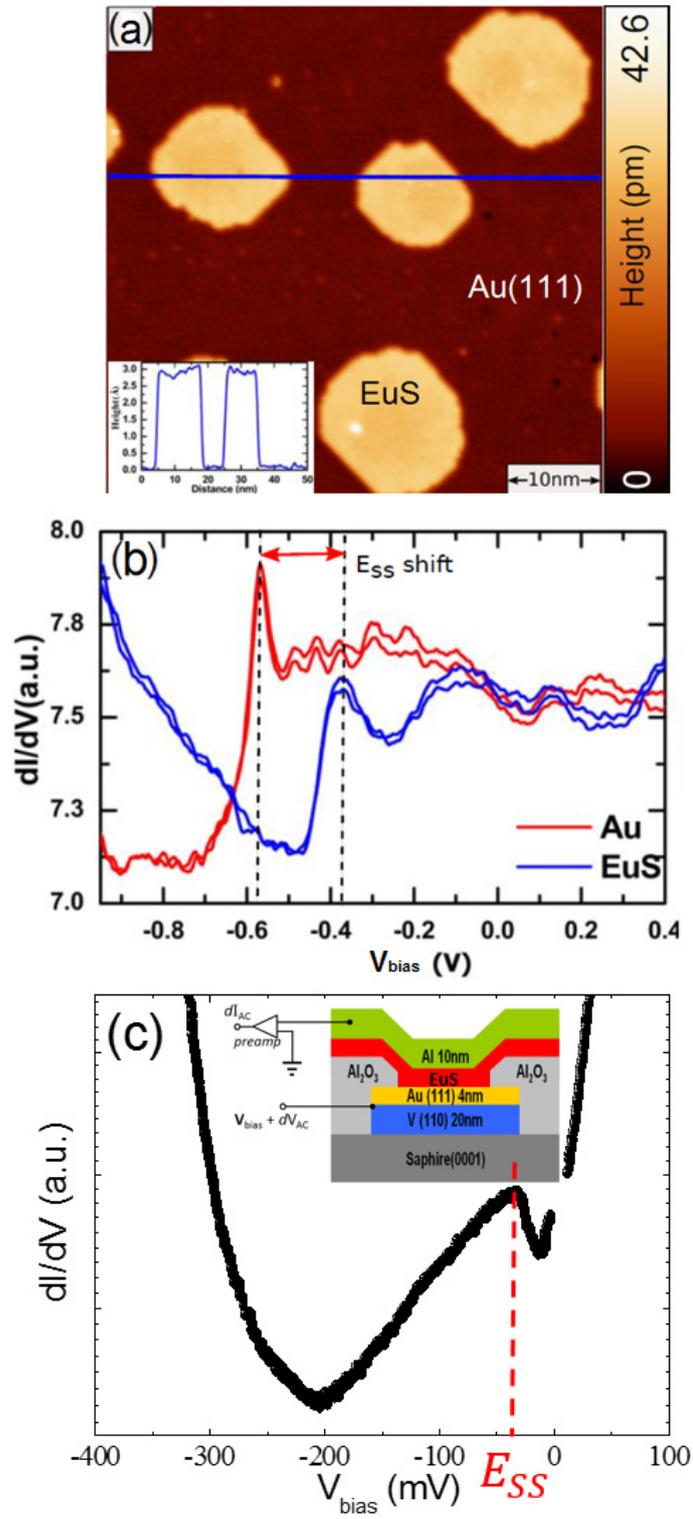

**Fig. 3**

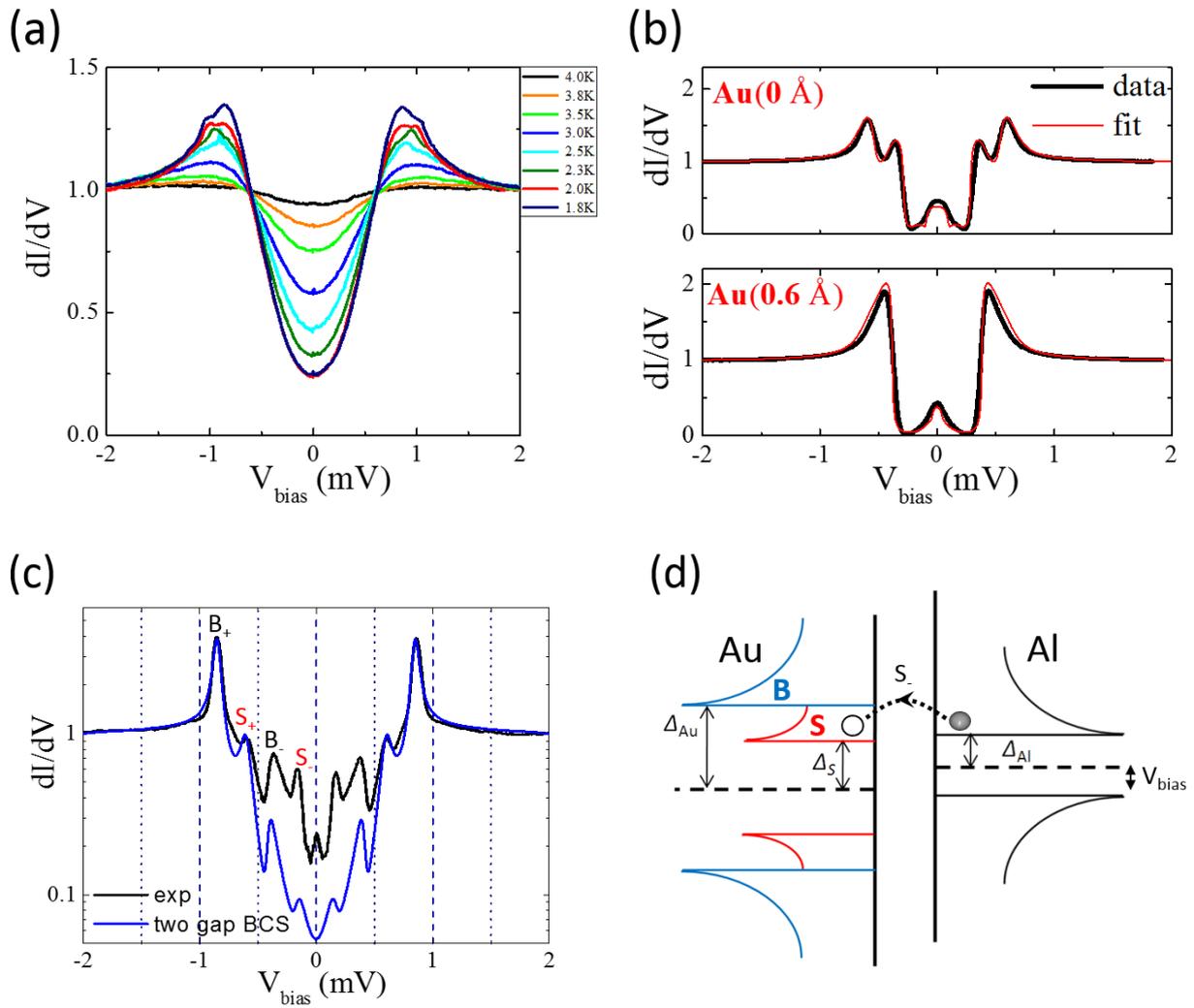

**Fig. 4**

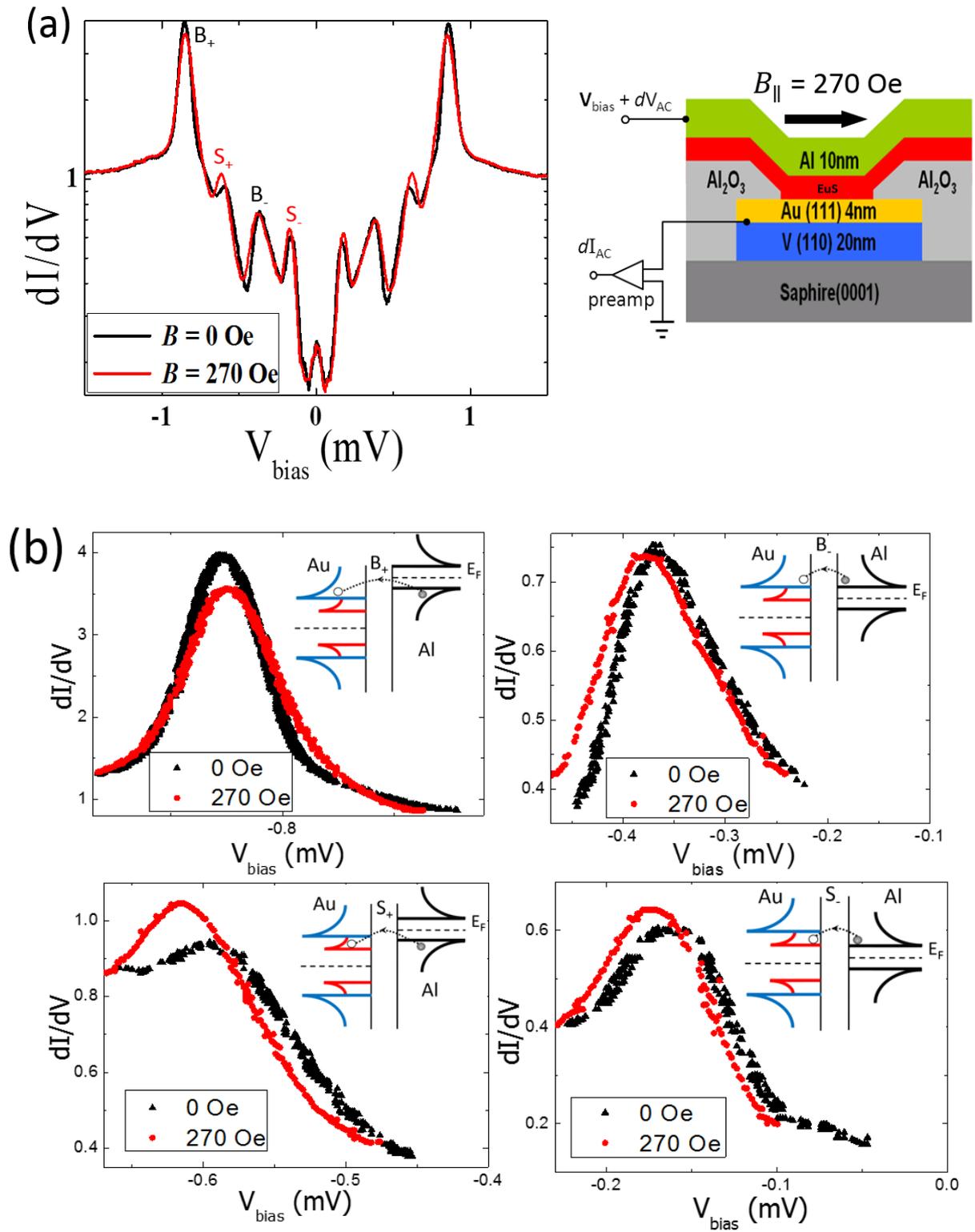

**Fig. 1: (a)** Large scale constant current STM topography image of 20nm thick V film grown on sapphire substrate. (Inset): schematic layout of the heterostructure. **(b)** High resolution STM image confirming V(110) surface.[4] **(c)** The large scale atomic terraces of 4nm Au grown on V. (Inset): dI/dV tunneling spectrum showing the edge of the surface energy band. **(d)** Atomic resolved STM image showing the hexagonal atomic lattice of Au(111).

**Fig. 2: (a)** STM topography image of sub-ML EuS grown on Au(111)/V surface. A continuous EuS layer is obtained when the thickness is above 3ML (~1nm), which causes difficulties for the STM scanning due to insulating EuS. The line scan profile (inset) shows the height of the EuS island is ~ 3Å. **(b)** STS spectrum obtained on top of bare Au(111) (red) and 1ML EuS island (blue) respectively. The bottom of the SS band shifts towards $E_F$ by ~ 0.2 eV. **(c)** With 2.4 nm EuS grown on Au(111), $E_F$ is found to be only ~ 34 meV above $E_{SS}$. Because 2.4 nm EuS is insulating, the $dI/dV$ spectra are measured in planar tunnel junction (TJ) devices (inset).

**Fig. 3: (a)** $dI/dV$ of the TJ shown in Fig. 2c inset (EuS thickness: 2.4 nm). Because $dI/dV$ depends on the DOS of the two layers (Au and Al) in the vicinity of the insulating barrier EuS, the $dI/dV$ gap reflects the induced SC in bulk Au(111). The doublet on the coherence peak is a signature of the SS gap. **(b)** A control sample showing that the doublet is not due to the MEF of EuS. The control TJs have one Al tunneling electrode coupled to EuS with decorated Au to tune the interface SOC (noted as Al/Al$_2$O$_3$/Au/Al/EuS). **(c)** $dI/dV$ of the same device as in **(a)** measured at T = 1.0 K (black curve). The multiple conductance peaks are well modeled (blue curve) by the S-I-S tunneling of the two band SC in Au (111) (Fig. 3d). **(d)** Schematics of the S-I-S tunneling model with Au having two induced SC gaps.

**Fig. 4: (a)** Tunneling conductance of planar TJ (schematic on the right) in the presence of a parallel magnetic field. Half of the tunneling peaks demonstrate contrasting responses to the magnetic field. **(b)** The zoomed-in data of each tunneling peak. For tunneling involving the bulk quasiparticles of Au, the conductance peak (red curves) is reduced by the applied magnetic field. For tunneling involving the SS quasiparticles of Au, the conductance peak (red curves) is enhanced by the applied magnetic field, which indicates the 2D nature of the quasiparticles in SS.

**Acknowledgments**:
P.W., S.M., P.L. and J.S.M. would like to acknowledge the support from John Templeton Foundation Grant No. 39944. P.W., S.M. and J.S.M. would like to acknowledge Office of Naval Research Grant N00014-13-1-0301 and N00014-16-1-2657 and National Science Foundation Grant DMR-1207469 and DMR-1700137.16

**References:**

1    Potter, A. C. & Lee, P. A. Topological superconductivity and Majorana fermions in metallic surface states. *Phys Rev B* **85**, 094516, (2012).

2    LaShell, S., McDougall, B. A. & Jensen, E. Spin Splitting of an Au(111) Surface State Band Observed with Angle Resolved Photoelectron Spectroscopy. *Phys Rev Lett* **77**, 3419, (1996).

3    Yan, B., Stadtmuller, B., Haag, N., Jakobs, S., Seidel, J., Jungkenn, D., Mathias, S., Cinchetti, M., Aeschlimann, M. & Felser, C. Topological states on the gold surface. *Nat Commun* **6**, (2015).

4    Wei, P., Katmis, F., Chang, C.-Z. & Moodera, J. S. Induced Superconductivity and Engineered Josephson Tunneling Devices in Epitaxial (111)-Oriented Gold/Vanadium Heterostructures. *Nano Lett* **16**, 2714-2719, (2016).

5    Hsieh, T. C. & Chiang, T. C. Spatial dependence and binding energy shift of surface states for epitaxial overlayers of Au on Ag(111) and Ag on Au(111). *Surf Sci* **166**, 554-560, (1986).

6    McMillan, W. L. Tunneling Model of the Superconducting Proximity Effect. *Phys Rev* **175**, 537-542, (1968).

7    Tomanic, T., Schackert, M., Wulfhekel, W., Sürgers, C. & Löhneysen, H. v. Two-band superconductivity of bulk and surface states in Ag thin films on Nb. *Phys Rev B* **94**, 220503, (2016).

8    Gupta, A. K., Crétinon, L., Moussy, N., Pannetier, B. & Courtois, H. Anomalous density of states in a metallic film in proximity with a superconductor. *Phys Rev B* **69**, 104514, (2004).

9    Wolz, M., Debuschewitz, C., Belzig, W. & Scheer, E. Evidence for attractive pair interaction in diffusive gold films deduced from studies of the superconducting proximity effect with aluminum. *Phys Rev B* **84**, 104516, (2011).

10   Hyart, T., van Heck, B., Fulga, I. C., Burrello, M., Akhmerov, A. R. & Beenakker, C. W. J. Flux-controlled quantum computation with Majorana fermions. *Phys Rev B* **88**, 035121, (2013).

11   Kliewer, J., Berndt, R., Chulkov, E. V., Silkin, V. M., Echenique, P. M. & Crampin, S. Dimensionality Effects in the Lifetime of Surface States. *Science* **288**, 1399-1402, (2000).

12   Aasen, D., Hell, M., Mishmash, R. V., Higginbotham, A., Danon, J., Leijnse, M., Jespersen, T. S., Folk, J. A., Marcus, C. M., Flensberg, K. & Alicea, J. Milestones Toward Majorana-Based Quantum Computing. *Physical Review X* **6**, 031016, (2016).

13   Mauger, A. & Godart, C. The magnetic, optical, and transport properties of representatives of a class of magnetic semiconductors: The europium chalcogenides. *Physics Reports* **141**, 51-176, (1986).

14   Smith, N. V. Phase analysis of image states and surface states associated with nearly-free-electron band gaps. *Phys Rev B* **32**, 3549-3555, (1985).

15   Pan, Y., Benedetti, S., Nilius, N. & Freund, H.-J. Change of the surface electronic structure of Au(111) by a monolayer MgO(001) film. *Phys Rev B* **84**, 075456, (2011).

16   Wei, P., Lee, S., Lemaitre, F., Pinel, L., Cutaia, D., Cha, W., Katmis, F., Zhu, Y., Heiman, D., Hone, J., Moodera, J. S. & Chen, C.-T. Strong interfacial exchange field in the graphene/EuS heterostructure. *Nat Mater* **15**, 711, (2016).

17   Katmis, F., Lauter, V., Nogueira, F. S., Assaf, B. A., Jamer, M. E., Wei, P., Satpati, B., Freeland, J. W., Eremin, I., Heiman, D., Jarillo-Herrero, P. & Moodera, J. S. A high-temperature ferromagnetic topological insulating phase by proximity coupling. *Nature* **533**, 513, (2016).

18   Wei, P., Katmis, F., Assaf, B. A., Steinberg, H., Jarillo-Herrero, P., Heiman, D. & Moodera, J. S. Exchange-Coupling-Induced Symmetry Breaking in Topological Insulators. *Phys Rev Lett* **110**, 186807, (2013).

19   Moodera, J. S., Santos, T. S. & Nagahama, T. The phenomena of spin-filter tunnelling. *J Phys-Condens Mat* **19**, 165202, (2007).

20   Nicolay, G., Reinert, F., Hufner, S. & Blaha, P. Spin-orbit splitting of the L-gap surface state on Au(111) and Ag(111). *Phys Rev B* **65**, 033407, (2001).

21   Hao, X., Moodera, J. S. & Meservey, R. Thin-film superconductor in an exchange field. *Phys Rev Lett* **67**, 1342-1345, (1991).

22   Li, B., Roschewsky, N., Assaf, B. A., Eich, M., Epstein-Martin, M., Heiman, D., Münzenberg, M. & Moodera, J. S. Superconducting Spin Switch with Infinite Magnetoresistance Induced by an Internal Exchange Field. *Phys Rev Lett* **110**, 097001, (2013).

23   Radebaugh, R. & Keesom, P. H. Low-Temperature Thermodynamic Properties of Vanadium. II. Mixed State. *Phys Rev* **149**, 217-231, (1966).

24   Meservey, R. & Tedrow, P. M. Spin-polarized electron tunneling. *Physics Reports* **238**, 173-243, (1994).

25   Maki, K. Pauli Paramagnetism and Superconducting State. II. *Prog Theor Phys* **32**, 29-36, (1964).

26   Fulde, P. & Maki, K. Theory of Superconductors Containing Magnetic Impurities. *Phys Rev* **141**, 275-280, (1966).

27   Meservey, R., Tedrow, P. M. & Fulde, P. Magnetic Field Splitting of the Quasiparticle States in Superconducting Aluminum Films. *Phys Rev Lett* **25**, 1270-1272, (1970).

28   Fulde, P. High field superconductivity in thin films. *Adv Phys* **22**, 667-719, (1973).

29   Dynes, R. C., Narayanamurti, V. & Garno, J. P. Direct Measurement of Quasiparticle-Lifetime Broadening in a Strong-Coupled Superconductor. *Phys Rev Lett* **41**, 1509-1512, (1978).

30   De Gennes, P. G. & Saint-James, D. Elementary excitations in the vicinity of a normal metal-superconducting metal contact. *Phys Lett* **4**, 151-152, (1963).

# Supplementary Information:

# Superconductivity in the surface state of noble metal gold and its Fermi level tuning by EuS dielectric

**Authors:** Peng Wei[1,3†*], Sujit Manna[1†*], Marius Eich[2,4], Patrick Lee[1*] and Jagadeesh Moodera[1,2*]

**Affiliations:**

1. Department of Physics, Massachusetts Institute of Technology, Cambridge, MA 02139, United States
2. Plasma Science and Fusion Center & Francis Bitter Magnet Laboratory, Massachusetts Institute of Technology, Cambridge, MA 02139, United States
3. Department of Physics and Astronomy, University of California, Riverside, CA 92521, United States
4. Solid State Physics Laboratory, ETH Zurich, 8093 Zurich, Switzerland

† These authors contributed equally to this work

\* Correspondence should be addressed to:

peng.wei@ucr.edu

smanna@mit.edu

palee@mit.edu

moodera@mit.edu

# S1: Suppression of Zeeman field induced quasiparticle spin splitting due to spin-orbit coupling (SOC)

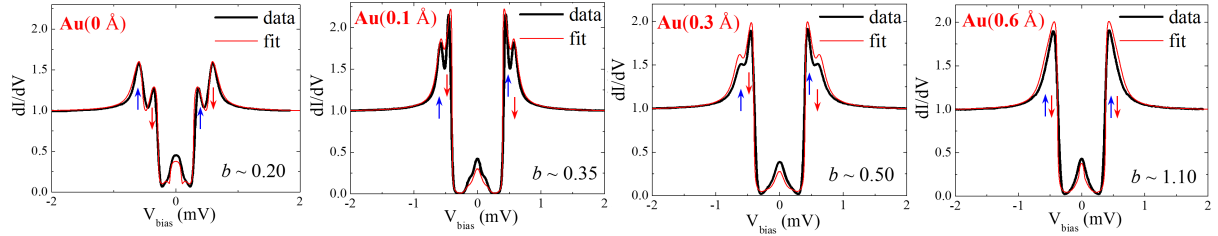

Figure S1 The measured $dI/dV$ spectra (black) of EuS/Al/Au/Al2O3/Al planar tunnel junctions with varying Au thickness. The Maki-Fulde model (red) is used to reproduce the experimental data (black), which demonstrates the increasing of the spin-orbit scattering parameter b when the Au thickness is increased. It can be clearly seen that the spin-splitting is completely suppressed when the Au layer thickness is only 0.6 Å.

The exchange field of an FMI (ferromagnetic insulator), such as EuS, is known to cause spin-splitting of the quasiparticle density of states (DOS) of a superconductor in proximity to the FMI.[1] When the energy of the spin-splitting is large enough compared to the superconductivity (SC) gap, Pauli limit is reached and SC is destroyed. In the presence of SOC in a superconductor, the spin is not a good quantum number. In this case the spin-splitting due to the exchange field is suppressed causing the increase of the upper critical field beyond the Pauli limit.[2] Such phenomena have been observed in Al thin films decorated with platinum (Pt) SOC scatters,[2,3] and in monolayer $NbSe_2$ with large Dresselhaus SOC.[4] We show that the SOC of Au is strong enough to completely suppress the spin splitting of quasiparticles in a s-wave superconductor.

We fabricated S-I-S tunnel junctions EuS/Al/Au/Al2O3/Al with $Al_2O_3$ acting as the tunnel barrier. A set of such tunnel junctions were fabricated with varying Au film thickness (from 0 Å to 0.6 Å), and simultaneously control junctions as well, enabled by our in-situ shadow mask technique. Moreover one of the Al layers was coupled to EuS so that the spin split quasiparticle DOS is generated by the proximity induced magnetic exchange coupling. The quasiparticle tunneling takes place between the two Al layers and the SOC strength is tuned by the thickness of Au.

Using Maki-Fulde model (see S3),[5-9] we simulate the $dI/dV$ tunneling spectra (Fig. S1) by tuning the dimensionless spin-orbit scattering rate $b = \hbar/3\Delta\tau_{so}$, where $\tau_{so}$ is the spin-orbit scattering time and $\Delta$ is the energy gap of the superconductor. The simulation results are compared with experimental data for a set of tunnel junctions (Fig. S1), where $b$ is systematically tuned by controlling the thicknesses of Au at the junction interface. In the absence of interfacial Au, (i.e., Au 0Å in Fig. S1), the spin splitting of the quasiparticle DOS is well-defined. The simulation using Maki-Fulde model suggests $b \sim 0.20$. We note that $b \sim 0.03$ is expected for Al with no spin-orbit scatters,[2] while the higher $b$ value we get could be attributed to the spin-flip scatterings (magnetic scattering), which adds extra terms to the Maki-Fulde formula and thus increases the value of $b$.[6,10] The magnetic scatterings could come from the unaligned magnetic $Eu^{2+}$ ions at the interface and domain walls in EuS at zero or low fields. However, the systematic increase of $b$ (Fig. S1) as Au is introduced at the interface is due to the increase of spin-orbit scattering. As the Au decoration is increased, the spin splitting of the quasiparticles becomes less prominent. The spin split $dI/dV$ peaks merge and is further accompanied by a significant decrease of the peak for the spin with higher energy (Fig. S1). When Au is 0.6 Å thick, much less than a monolayer, the spin splitting is completely suppressed (Fig. S1) with $b = \hbar/3\Delta\tau_{so} \sim 1.10$. Although our control experiments are carried out in Al junctions, the spin orbit scattering rate of Au is not expected to depend on the superconductor material. Therefore, with 4 nm thick Au at the tunnel barrier interface, we do not expect to see the spin splitting of quasiparticles in the $dI/dV$ spectrum of Al/EuS/Au(111)/V tunnel junctions.

**S2: The fittings of SC tunneling in the presence of Zeeman splitting, spin-orbit coupling and surface gap**

The DOS of a superconductor under an external magnetic field and SOC is described by Maki-Fulde theory,[5-8,11] which takes into account the orbital depairing, Zeeman splitting and spin-orbit scattering effects. Two additional corrections describing the life-time of the quasiparticles are introduced to the energy term in the BCS model as:

$$u_\pm = \frac{E \mp \mu_B H}{\Delta} + \zeta \frac{u_\pm}{(1-u_\pm^2)^{1/2}} + b \frac{u_\pm - u_\mp}{(1-u_\mp^2)^{1/2}} \quad (S1)$$

Here $\zeta$ is the orbital depairing parameter, which describes the life-time of Cooper pair breaking. $b$ is the spin-orbit scattering parameter as $b = \hbar/3\Delta\tau_{so}$, which describes the life-time of spin flip scattering due to SOC.[2] $\mu_B H$ is the Zeeman energy term that a quasiparticle would gain in the presence of a magnetic field applied parallel to the thin film superconductor. The DOS of the spin-up and spin-down quasiparticles are given as:

$$\mathcal{N}_{\uparrow\downarrow}(E) = \frac{\mathcal{N}(0)}{2} sgn(E) Re\left[\frac{u_\pm}{(u_\pm^2 - 1)^{1/2}}\right] \quad (S2)$$

$\mathcal{N}(0)$ is the normal state DOS of the quasiparticles at the Fermi level. The indices + (-) corresponds to the spin ↑ (↓). As is pointed previously, we solve $u_\pm$ in Eq. S1 numerically using Newton-Raphson methods in a recursive manner.[9,11,12] Thus, the tunneling current is:

$$I(V) \propto \int (\mathcal{N}_\uparrow(E) + \mathcal{N}_\downarrow(E))\mathcal{N}_{Al}(E - eV)[f(E - eV, T) - f(E, T)] dE \quad (S3)$$

Here $\mathcal{N}_{Al}(E)$ is the quasiparticle DOS of aluminum (top tunneling electrode) and $f(E, T)$ describes the Fermi-Dirac distribution. The tunneling conductance $dI/dV$ is simulated according to Eq. S1-S3 and is compared to our measured tunnel conductance so as to extract the spin-orbit parameter $b$ (Fig. S1 and Fig. 3b). The strength of the SOC due to Au, given by $b$, is comparable to that due to the Pt spin-orbit scatters as reported previously.[13]

1   Hao, X., Moodera, J. S. & Meservey, R. Thin-film superconductor in an exchange field. *Phys Rev Lett* **67**, 1342-1345 (1991).
2   Meservey, R. & Tedrow, P. M. Spin-polarized electron tunneling. *Physics Reports* **238**, 173-243 (1994).
3   Tedrow, P. M. & Meservey, R. Experimental Test of the Theory of High-Field Superconductivity. *Phys Rev Lett* **43**, 384-387 (1979).
4   Xi, X., Wang, Z., Zhao, W., Park, J.-H., Law, K. T., Berger, H., Forro, L., Shan, J. & Mak, K. F. Ising pairing in superconducting NbSe2 atomic layers. *Nat Phys* **12**, 139-143 (2016).
5   Maki, K. Pauli Paramagnetism and Superconducting State. II. *Prog Theor Phys* **32**, 29-36 (1964).
6   Fulde, P. & Maki, K. Theory of Superconductors Containing Magnetic Impurities. *Phys Rev* **141**, 275-280 (1966).
7   Meservey, R., Tedrow, P. M. & Fulde, P. Magnetic Field Splitting of the Quasiparticle States in Superconducting Aluminum Films. *Phys Rev Lett* **25**, 1270-1272 (1970).
8   Fulde, P. High field superconductivity in thin films. *Adv Phys* **22**, 667-719 (1973).
9   Alexander, J. A. X., Orlando, T. P., Rainer, D. & Tedrow, P. M. Theory of Fermi-liquid effects in high-field tunneling. *Phys Rev B* **31**, 5811-5825 (1985).
10  Bruno, R. C. & Schwartz, B. B. Magnetic Field Splitting of the Density of States of Thin Superconductors. *Phys Rev B* **8**, 3161-3178 (1973).
11  Meservey, R., Tedrow, P. M. & Bruno, R. C. Tunneling measurements on spin-paired superconductors with spin-orbit scattering. *Phys Rev B* **11**, 4224-4235 (1975).
12  Worledge, D. C. & Geballe, T. H. Maki analysis of spin-polarized tunneling in an oxide ferromagnet. *Phys Rev B* **62**, 447-451 (2000).
13  Tedrow, P. M. & Meservey, R. Critical magnetic field of very thin superconducting aluminum films. *Phys Rev B* **25**, 171-178 (1982).